\documentclass[useAMS,usenatbib]{mnras}

\usepackage{amsmath}
\usepackage{amssymb}
\usepackage{times,txfonts}
\usepackage{graphicx}

\usepackage{newtxtext,newtxmath}
\usepackage[T1]{fontenc}
\usepackage{ae,aecompl}
\usepackage{float}

\newcommand{\ud}{\ensuremath{\mathrm{d}}}

\title[Fallback and Spin-Kick Alignment]{Fallback onto Kicked Neutron Stars and its Effect on Spin-Kick Alignment}

\author[M\"uller]{
Bernhard M\"uller$^{1,2}$\thanks{E-mail: bernhard.mueller@monash.edu}\\
$^{1}$School of Physics and Astronomy, Monash University, VIC 3800, Australia\\
$^{2}$ ARC Centre of Excellence for Gravitational Wave Discovery -- OzGrav
}

\begin{document}

\label{firstpage}
\pagerange{\pageref{firstpage}--\pageref{lastpage}}

\maketitle
             
\begin{abstract}
Fallback in core-collapse supernova explosions is potentially of significant importance for the birth spins of neutron stars and black holes. It has recently been pointed out that the angular momentum imparted onto a compact remnant by fallback material is subtly intertwined with its kick because fallback onto a moving neutron star or black hole will preferentially come for a conical region around its direction of travel. We show that contrary to earlier expectations such one-sided fallback accretion onto a neutron star will tend to produce spin-kick misalignment. Since the baroclinic driving term in the vorticity equation is perpendicular to the nearly radial pressure gradient, convective eddies in the progenitor as well as Rayleigh-Taylor plumes growing during the explosion primarily carry angular momentum perpendicular to the radial direction. Fallback material from the accretion volume of a moving neutron star therefore carries substantial angular momentum perpendicular to the kick velocity. We estimate the seed angular momentum fluctuations from convective motions in core-collapse supernova progenitors and argue that accreted fallback material will almost invariably be accreted with the maximum permissible specific angular momentum for reaching the Alfv\'en radius. This imposes a limit of $\mathord{\sim}10^{-2}M_\odot$ of fallback accretion for fast-spinning young neutron stars with periods of $\mathord{\sim}20\,\mathrm{ms}$ and less for longer birth spin periods.
\end{abstract}


\section{Introduction}
\label{sec:intro}
One of the key challenges in the theory of core-collapse supernova explosions of massive stars is to explain the birth properties of neutron stars and black holes. Phenomenological supernova models \citep{ugliano_12,sukhbold_16,mueller_16a,ebinger_19,ertl_20} and multi-dimensional simulations \citep[for recent reviews, see][]{mueller_20,burrows_21}
have already shed considerable light on the stellar progenitor properties that determine whether the collapse results in neutron star or black hole formation, and on the birth mass 
distribution of compact objects. The theoretical understanding of neutron star birth velocities (kicks) now appears very mature as well and points to the ``gravitational tugboat 
mechanism'' in asymmetric explosions as a natural way to account for neutron star kicks of several $100\, \mathrm{km}\, \mathrm{s}^{-1}$ and up to
$\gtrsim 1000 \, \mathrm{km}\, \mathrm{s}^{-1}$
\citep{scheck_06,wongwathanarat_10b,wongwathanarat_13,mueller_17,mueller_19a,bollig_21,coleman_22}
as observed in nature, although asymmetric neutrino emission \citep{stockinger_20,coleman_22} will also play a -- probably subdominant -- role, and other, more hypothetical kick mechanisms are conceivable. 
It has also been realised that the tugboat
mechanism may  produce black holes with
kicks in ``fallback explosions'' with partial
mass ejection \citep{janka_13,chan_18,chan_20b}, in line with observational evidence.

The origin of neutron star spins remains more elusive for several reasons. On the one hand, it is far from clear to what extent the spins are determined by the angular momentum of the progenitor core, or rather by spin-up and spin-down processes during the explosion. Recent multi-dimensional simulations suggest that spin-up by asymmetric accretion could be the dominant mechanism for setting neutron star birth spins and
might produce a realistic range of birth spin rates from a few seconds down to $\mathord{\sim}10\, \mathrm{s}$ as in the observed pulsar population \citep{popov_12,igoshev_13,noutsos_13}. However, our incomplete knowledge
of the interior rotation rates of massive stars
\citep{heger_00,heger_05,langer_12,fuller_19} and uncertainties about early spin-down
processes mediated by magnetic fields make it difficult to draw firm conclusions.

Among compact object birth properties, perhaps the greatest challenge to current core-collapse supernova models consists in strong observational evidence that neutron star spins and kicks tend to be aligned \citep{kramer_03,johnston_05,noutsos_12,noutsos_13,yao_21}. 
Several mechanisms for spin-kick alignment have been proposed over the years; some of these are purely hydrodynamic in nature, while others invoke kicks generated by asymmetric emission of neutrinos or radiation from very strongly magnetised neutron stars
\citep[e.g.,][]{harrison_75,spruit_98,arras_99b,lai_01,socrates_05,wang_07,fragione_23}.
Multi-dimensional simulations have rather stubbornly refused to bear out any of the suggested hydrodynamic mechanisms and cannot yet reproduce systematic spin-kick alignment
\citep{wongwathanarat_13,powell_19,powell_20,janka_22,powell_23}. Some recent three-dimensional models by \citet{coleman_22} show strong
spin-kick alignment or anti-alignment, but the majority of their models still exhibit large angles between the spin and kick direction.

\begin{figure}
    \centering
    \includegraphics[width=\linewidth]{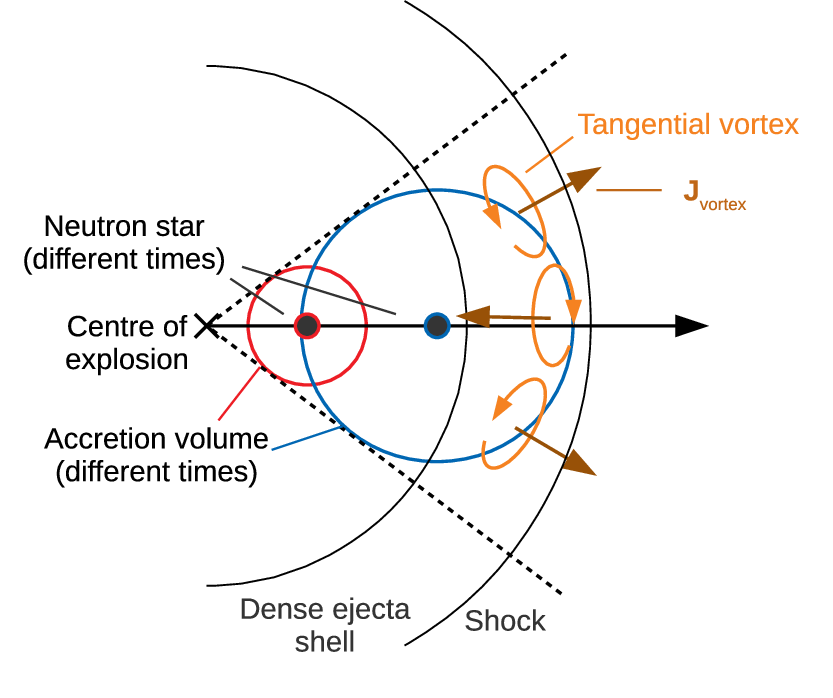}
    \caption{Simplified sketch of one-sided neutron star fallback accretion as outlined by \citet{janka_22}. As
    the neutron star moves away from the centre of explosion, so does the spherical accretion volume (red and blue circles), tracing out a conical region within the ejecta from which fallback material may be supplied. The ejecta are compressed into a dense shell behind the shock (outer black circle). \citet{janka_22} posit that vortices originating from convective motions in the progenitor star are ``flattened'' as they are run over and radially compressed by the shock so that their vorticity and angular momentum vectors $\mathbf{J}_\mathrm{vortex}$ predominantly point radially outward or inward.}
    \label{fig:vortex_janka}
\end{figure}

For this reason, \citet{janka_22} argued that spin-kick alignment could be due to fallback on longer time scales than the first $\mathord{\sim}1 \,\mathrm{s}$ that can presently be studied by rigorous three-dimensional simulations. They note that the neutron star will preferentially accrete fallback material along its direction of motion, and argue that stochastic vortical motions in the accreted material, ultimately arising from convective eddies in the outer shells, will then tend to magnify the neutron star angular momentum component along its direction of motion (Figure~\ref{fig:vortex_janka}).
Interestingly, the proposed mechanism does not require any progenitor rotation. The dynamical impact of angular momentum fluctuations from convective regions in fallback supernovae is also of broader interest because since it could lead to disk formation during the infall and power late-time outflows
\citep{gilkis_14,antoni_22,antoni_23,soker_23}.

In this paper, we shall revisit this idea and argue that this mechanism is more likely to destroy spin-kick alignment than to facilitate it if the interaction of the supernova shock wave with convective eddies in the progenitor is examined more closely. If true this gives a different twist to the original idea of \citet{janka_22}. However, as the phenomenon of one-sided accretion pointed out by \citet{janka_22} is likely a robust feature of fallback explosion, their idea can then  be used to place constraints on the amount of fallback in supernova explosions. Based on the typical convective velocities in supernova progenitors and physical constraints on the accretion process, we shall discuss implications for the permitted amount of fallback in typical core-collapse supernovae.

\section{Summary of Model for Spin-Kick Alignment}
It is useful to first review the geometrical picture
for fallback on a moving neutron star outlined by \citet{janka_22}. Their proposed scenario is sketched in
Figure~\ref{fig:vortex_janka}.
\citet{janka_22} point out that fallback is likely to affect ejecta
that move with sufficiently small velocity $v$ with respect to a neutron star of mass $M$ for their kinetic energy to be of the same order to their potential energy or smaller. This defines the
maximum distance of prospective fallback material to
the neutron star $R_\mathrm{acc}$ \citep{shapiro_83} (accretion radius),
\begin{equation}
\label{eq:racc}
    \xi \frac{G M}{R_\mathrm{acc}}
    = \frac{1}{2}v^2,
\end{equation}
with a non-dimensional factor $\xi$ of order unity. \citet{janka_22}
then demonstrate that for a homologously expanding explosion and a few further assumptions, the
volume defined by Equation~(\ref{eq:racc}) is indeed roughly spherical and centred around the current location of the neutron star, and that the distance of the neutron star from the geometric centre of the explosion will easily be comparable to or larger than $R_\mathrm{acc}$. This implies very asymmetric accretion by the neutron star, mostly from ahead of its direction of travel, where the accretion volume reaches furthest (or is most likely to reach)  into the shell of dense, shock-compressed ejecta.

For such asymmetric fallback of material forward of the neutron star to lead to spin-kick alignment, \citet{janka_22} make the
crucial assumption that vortex motions in the fallback material
primarily occur perpendicular to the radial direction such
that the vorticity $\mathbf{\omega}$ and specific angular momentum
$\mathbf{j}$ of the accreted material has a dominant \emph{radial}
component. Under the assumption of \citet{janka_22} that the radial
ejecta velocities $v_r$ are perfectly spherically symmetric, that only non-radial (transverse) velocity perturbations $\mathbf{v}_t$ are present in the ejecta, and that no further exchange of angular momentum occurs during fallback, the angular momentum of accreted vortices is simply
\begin{equation}
    \mathbf{J}=\int \rho \mathbf{r} \times \delta \mathbf{v}_t\,\ud V,
\end{equation}
where the integral extends over the fallback region.\footnote{See Section~4.2 in \citet{janka_22} for a detailed discussion why other terms cancel out, and Section~4.2.3 for the case with density inhomogenities, but still with a perfectly spherical radial velocity field $v_r(r)$.}
Even if the neutron star may accrete vortices with a 
stochastically varying sense of circulation, the net effect will be 
to add a large amount of angular momentum with a direction close
to the neutron star displacement vector $\mathbf{D}_\mathrm{NS}$ from the centre of explosion
and the neutron star velocity $\mathbf{v}_\mathrm{NS}$. If the angular momentum added by fallback is larger in terms of magnitude than the initial neutron star angular momentum, this would drive the neutron star towards spin-kick alignment.

\begin{figure*}
    \centering
    \includegraphics[width=0.85\linewidth]{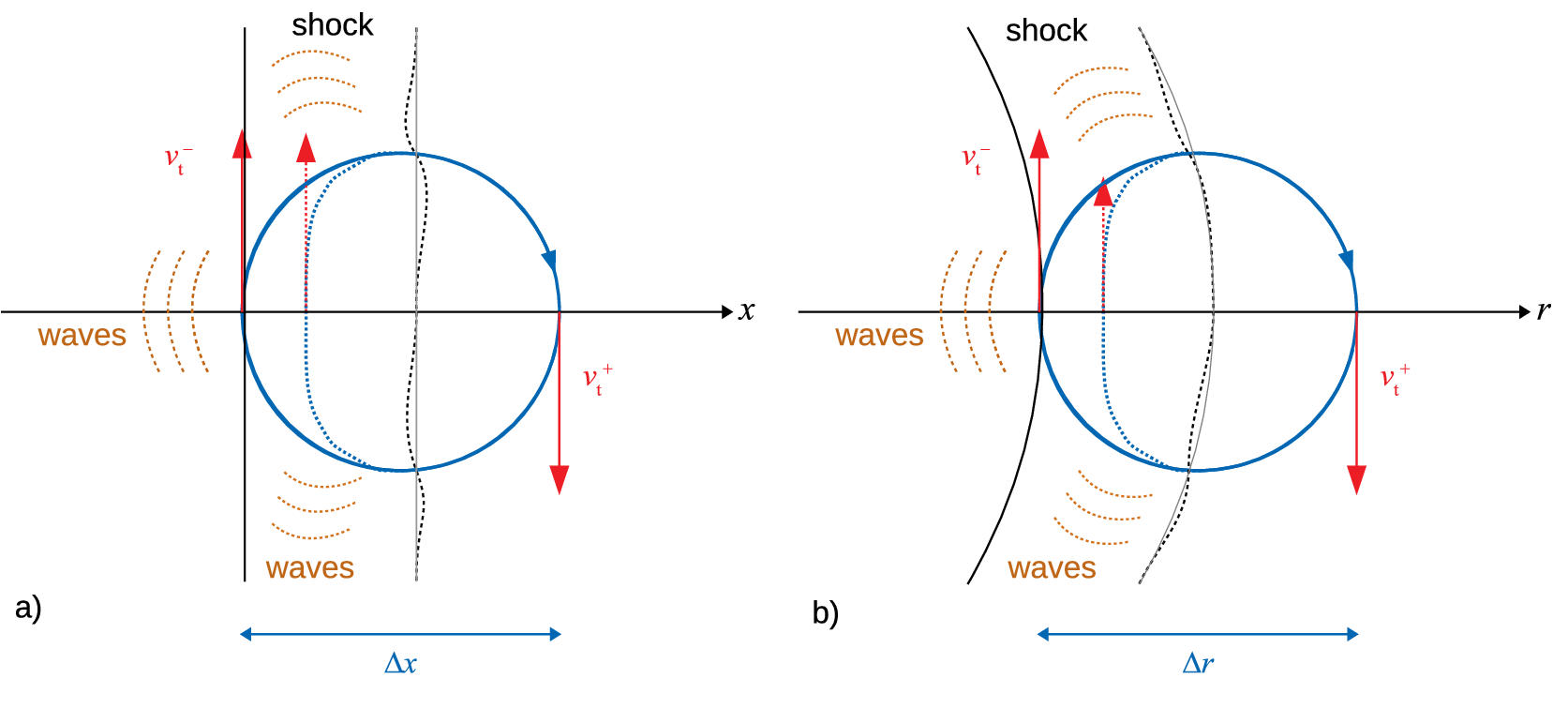}
    \caption{Interaction of a convective vortex a) with a planar shock wave and b) with a spherical shock wave. For the sake of simplicity, the pre-shock medium is assumed to have homogeneous density so that a circular flow pattern 
    with constant velocity along the vortex fulfils the solenoidal condition for anelastic flow. The shock and the structure of the vortex are depicted at two stages, i.e., when the shock has just hit the eddy (solid lines) and when it has run further across half of the eddy (dashed lines). Red arrows indicate the transverse velocity
    $v_\mathrm{t}^-$ furthest to left (furthest inward) of the eddy and $v_\mathrm{t}^+$ furthest to the right (furthest outside). 
    The direction of shock propagation (in the $x$-direction and in the radial direction, respectively) and the initial extent $\Delta x$ and $\Delta r$ of the eddy are  
    also shown. In the planar case $v_\mathrm{t}^-$ remains unchanged as the shock stars to hit the eddy and is not considerably affected later, except that waves behind the shock (orange)  transport transverse momentum around to some degree. The planar shock is also slightly corrugated due to slightly inhomogeneous pre-shock conditions and wave activity downstream. In the case of a spherical shock, the picture is similar, except for one important difference. As the inner par of the eddy moves outwards after being shocked, the transverse velocity $v_\mathrm{t}^-$ decreases due to fictitious forces (which ensure specific angular momentum conservation). Thus the total angular momentum of the shocked eddy remains conserved despite radial compression.
    }
    \label{fig:eddy_shock}
\end{figure*}

The critical question for this mechanism is whether the specific angular momentum of vortices in the compressed ejecta shell indeed tends to be aligned with the radial direction. There are two possible (and interrelated) scenarios for the formation of such vortices. The vortices may simply be relics of convective eddies from the pre-collapse stage that are compressed as they are run over by the shock \citep{janka_22}. The desired vortices may also be generated by Rayleigh-Taylor 
and Richtmeyer-Meshkov instabilities \citep{richtmyer_60,zhou_17a,zhou_17b}
that operate behind the supernova shock \citep[e.g.,][]{chevalier_76,mueller_91,fryxell_91,kifonidis_06,wongwathanarat_15}, with the seeds of the instabilities either provided by the aforementioned eddies or initial explosion asymmetries due to convection \citep{herant_92,herant_94,burrows_95,janka_95,janka_96}
or the standing-accretion shock instability \citep{blondin_03} in the supernova core. \citet{janka_22} argues that for strictly
spherical expansion with $v_r=v_r(r)$, vortical motions in the 
ejecta must inevitably be purely transverse. Indeed, it may appear at first glance that ``closed'' vortices with transverse vorticity and angular momentum require departures from a spherically symmetric expansion law (Figure~\ref{fig:vortex_janka}).

\section{Evolution of Vortices during the Supernova Explosion}
However, these arguments do not appear conclusive and suffer from an inconsistency in the assumptions. To elucidate the issue, let us first consider the case of a convective eddy hit by the supernova shock and also make the assumption that the density and pressure fields are perfectly spherical. The eddy
will be squeezed due to radial compression, but if $\nabla P$
and the gravitational acceleration $\mathbf{g}$ are strictly radial, this implies that the specific angular momentum $\mathbf{j}=\mathbf{r} \times \mathbf{v}$ in the shocked ejecta does not change,
\begin{equation}
\frac{\ud \mathbf{j}}{\ud t} =- \mathbf{r}\times \frac{\nabla P}{\rho}+\mathbf{r}\times \mathbf{g}=0,
\end{equation}
and hence the total angular momentum in the squeezed vortex
volume $V(t)$ should also be constant
\begin{align}
    \frac{\ud }{\ud t}
    \int_{V(t)} \rho \mathbf{j} \,\ud V=0,
\end{align}
and identical to the angular momentum of the vortex before collapse.
Without invoking non-spherical pressure perturbations, there is no mechanism for tilting eddies into a different plane and changing the direction of their angular momentum. Since convective motions in the pre-collapse phase innately involve upward and downward radial motions across the largest available scale (the width of the convection zone), and lateral flow to ``close the loop'' between them at the top and bottom of the convection zone, it is intuitively clear that the angular momentum of the large-scale eddies will generally be dominated by the non-radial components. More formally,
one can consider the vorticity equation,
\begin{equation}
\label{eq:vorticity}
    \frac{\ud \boldsymbol{\omega}}{\ud t}
    =
    (\boldsymbol{\omega}\cdot\nabla) \mathbf{v}-
    \    (\boldsymbol{\omega}\cdot\nabla) \mathbf{v}-{\omega} (\nabla \cdot\mathbf{v})
    +\frac{\nabla\rho \times \nabla P}{\rho^2},
\end{equation}
and note that the baroclinic  term $\nabla \rho \times \nabla P/\rho^2$ is the driving term for the vorticity evolution in convective motions\footnote{Note that to effectively capture a temporally-averaged steady state of convection, one would need to perform Reynolds averaging on Equation~(\ref{eq:vorticity}) and apply appropriate closures, but this is beyond the scope of this paper; a qualitative consideration of the driving term is sufficient for our purpose.} (as it corresponds to the buoyancy term as the driving term in the momentum equation). Since the non-radial components $\nabla_\mathrm{t} \rho$ and $\nabla_\mathrm{t} P$ of the density and
pressure gradients are small compared to the radial components
$\nabla_\mathrm{r} \rho$ and $\nabla_\mathrm{r} P$, we have
\begin{equation}
    \frac{\nabla\rho \times \nabla P}{\rho^2}
    \approx
    \frac{\nabla_\mathrm{t} \delta \rho \times \nabla_\mathrm{r} P+ \nabla_\mathrm{r} \delta \rho \times \nabla_\mathrm{t} P}{\rho^2} \perp \mathbf{r},
\end{equation}
i.e., for small density and pressure perturbations (as encountered in subsonic convection in supernova progenitors), buoyancy generates non-radial vorticity perturbations. Incidentally,
the same argument also precludes the generation of radial vorticity
by Rayleigh-Taylor instability behind
the supernova shock as long as the pressure field is assumed to be spherically symmetric, as this instability also hinges on the baroclinic term \citep{zhou_17a}. In a nearly spherical explosion, Rayleigh-Taylor instability will again primarily generate non-radial vorticity. As long as no extreme asymmetries are assumed in the progenitor star or the supernova explosion, it appears hard to avoid the presence of vortices with \emph{non-radial} vorticity in the shocked ejecta.

\begin{figure*}
    \centering
    \includegraphics[width=0.33\linewidth]{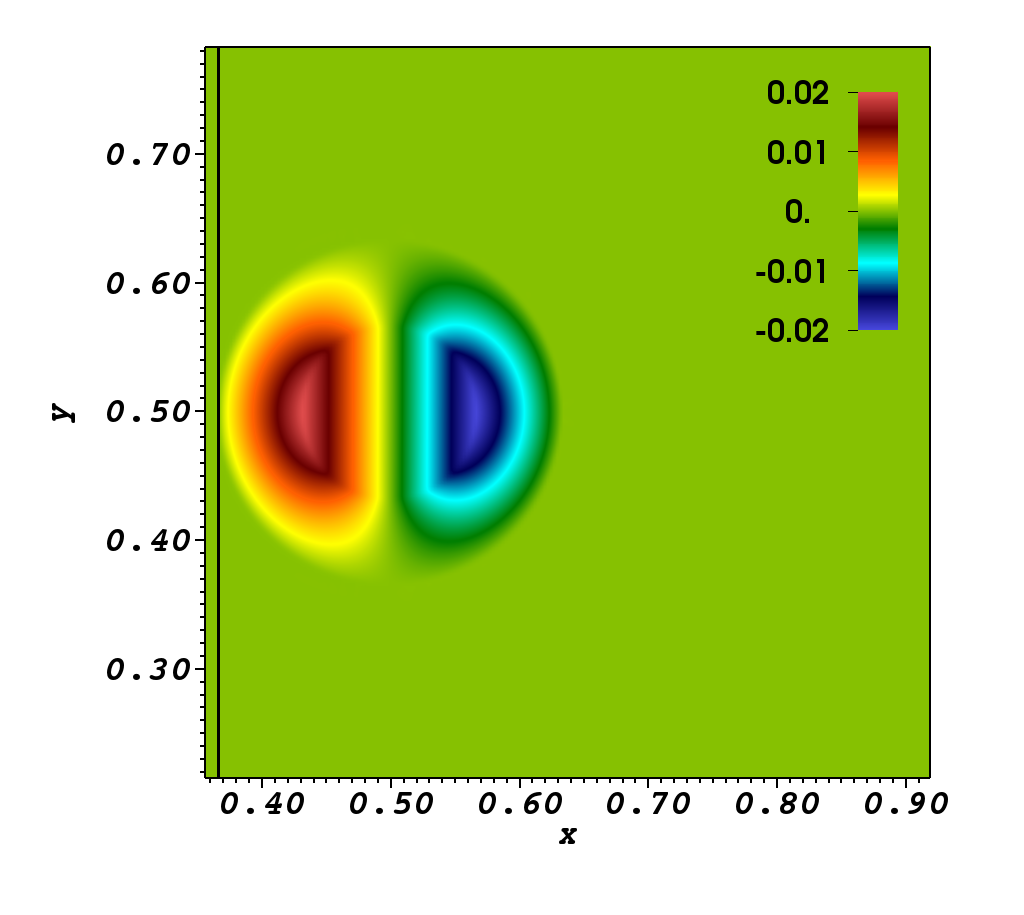}
    \includegraphics[width=0.33\linewidth]{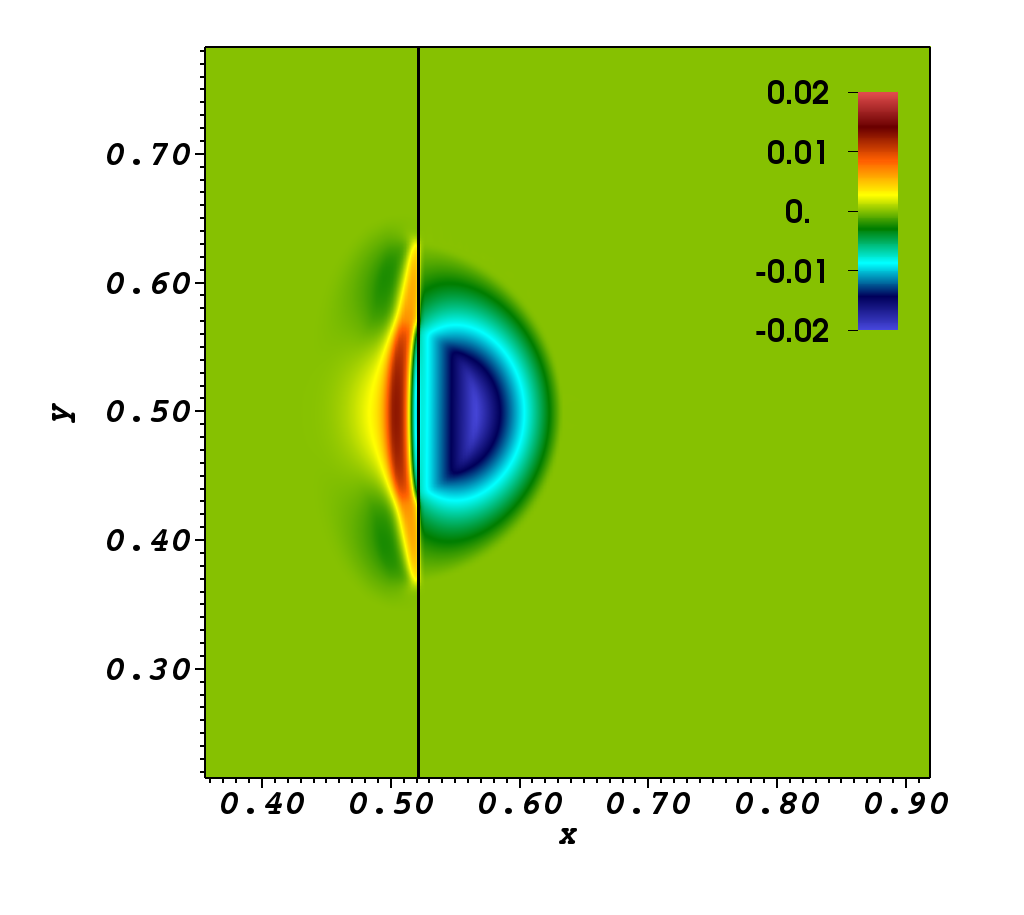}
    \includegraphics[width=0.33\linewidth]{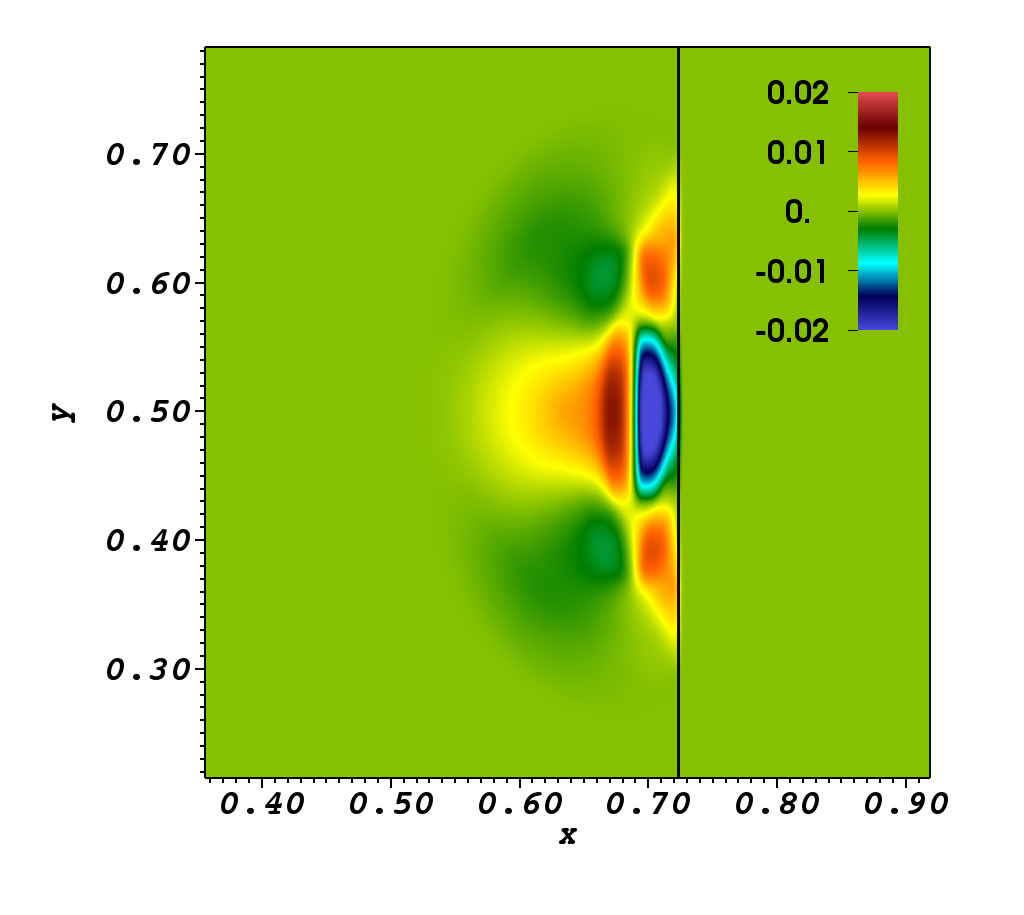}
    \caption{Transverse velocity (in dimensionless units) for a vortex overrun by a shock (black solid line) before (left panel), during (middle panel), and after the passage of the shock (right panel). The vortex is flattened to a more oblate shape, and in addition, waves that propagate sideways and downstream of the shock redistribute transverse momentum. The maximum and minimum transverse velocities before and after the passage of the shock are similar.}
    \label{fig:2dsim}
\end{figure*}

To somewhat reconcile these arguments with the intuitive notion of ``vortex flattening'' by shock compression, it is useful
to consider the interaction of the supernova shock with existing vortices in more depth (Figure~\ref{fig:eddy_shock}). The problem of shock-vortex interaction has been studied more formally using analytic theory \citep{wouchuk_07,wouchuk_09} and simulations
\citep{ellzey_95,barbosa_01,zhang_05} in the case of a planar shock,
and also for a stalled supernova shock during the pre-explosion phase \citep{huete_18}, but a qualitative analysis will suffice here.
If we assume
the curvature of the shock is negligible over the scale of a vortex,
the transverse velocity perturbations will simply be conserved across the shock. Upstream transverse velocity perturbations will leave velocity perturbations downstream of the shock, and also corrugate the shock surface. The corrugation implies that the shocked material \emph{will} experience some change in vorticity, but while the vortex experiences some squeezing, it is not destroyed. However, the downstream perturbation will no longer be purely vortical, and acoustic waves will be produced by the shock-vortex interaction. Further coupling between vortical and acoustic perturbations will occur as the vortex expands in the wake of the shock \citep[cp.\ the analogous process during the collapse phase,][]{abdikamalov_16,abdikamalov_19}.

To further illustrate this phenomenon, we simulate the interaction of a planar shock with a vortex in two dimensions. We set up
a non-dimensional problem on a quadratic domain of size $\Delta x=\Delta y=1$. A shock perpendicular to the $x$-direction is set up
at $x=0.25$ with post-shock density $\rho_\mathrm{p}=1.32$, pressure  $P_\mathrm{p}=16.77$, and velocity $v_\mathrm{p}=8.48$
and pre-shock density $\rho_0=0.2$ and pressure $P_0=0.2$; a perfect gas with $\gamma=4/3$ is assumed. The pre-shock medium is initially at rest except for a vortical
velocity perturbation 
$\mathbf{\delta v} =v_\mathrm{rot} = 
 \frac{\mathbf{x}-\mathbf{x}_0}{|\mathbf{x}-\mathbf{x}_0|} \times\mathbf{e}_z$
centred around $\mathbf{x}_0=(0.5,0)$. The rotational
velocity $v_\mathrm{rot}$ is chosen in close analogy to
the Gresho vortex \citep{gresho_90},
\begin{equation}
    v_\mathrm{rot}= 
    \left\{
    \begin{array}{ll}
    0.3 |\mathbf{x}-\mathbf{x}_0|, & |\mathbf{x}-\mathbf{x}_0|<1/15 \\
    0.04 - 0.3 |\mathbf{x}-\mathbf{x}_0| & 1/15\leq |\mathbf{x}-\mathbf{x}_0|<1/15 < 2/15\\
    0, \mathrm{else}
    \end{array}
    \right.
\end{equation}
although no pressure perturbations are imposed. The equations of compressible hydrodynamics are solved with a finite-volume
code that employs second order reconstruction with the van Leer limiter \citep{van_leer_74}, the HLLC Riemann solver \citep{toro_94} with an adaptive switch for the more dissipative HLLE solver
\citep{einfeldt_88} at shocks to avoid odd-even decoupling
and the carbuncle phenomenon \citep{quirk_94}, and second-order
Runge-Kutta time stepping. Transverse velocities at different
stages of the shock-vortex interaction are shown in Figure~\ref{fig:2dsim}.
For this particular problem, the distortion of the shock is minimal, but deviations from the initial irrotational flow pattern are sufficient to launch  waves transverse to the shock and somewhat affect the transverse velocities in the shocked eddy. However, the vortex clearly survives with transverse velocities
similar to the initial conditions.

It might still appear that if the transverse velocities in a vortex
are almost conserved as the vortex is shocked, the net angular
momentum of the vortex could decrease simply by compression.
As the total angular momentum $J$  of the vortex is approximately
\begin{align}
    J&=\int r \rho v_\mathrm{t}\,\ud V
    \approx 
    J \Delta M/2\, [(r+\Delta r/2) v_\mathrm{t}^+ + (r-\Delta r/2) v_\mathrm{t}^-] 
    \nonumber
    \\
    &=\Delta M/2\, \Delta r \, (v_\mathrm{t}^+ + v_\mathrm{t}^-),
    \label{eq:j_eddy}
\end{align}
in terms of the mass $\Delta M$ of the vortex, its radial extent $\Delta r$, and the transverse velocities $v_\mathrm{t}^+$ and $v_\mathrm{t}^-$
in the outer and inner portion of the loop. Thus, if $v_\mathrm{t}^+$ and $v_\mathrm{t}^-$ are unaffected by the shock, compression of $\Delta r$
might appear to reduce $J$. However, it must be borne in mind that the vortex interacts with a spherical shock wave
(Figure~\ref{fig:eddy_shock}b). In this case, the magnitude of the transverse velocity $v_\mathrm{t}^-$ in the inner portion of the vortex will decrease due to
fictitious forces that maintain angular momentum conservation (neglecting transverse pressure gradients) while the radial extent $\Delta r$ of the vortex shrinks as the shock runs across it. If the subtle impact of fictitious forces on the transverse velocities $v_\mathrm{t}$ is taken into account during shock-vortex interaction, the total angular momentum of the vortex will be conserved as expected.

In summary, it seems improbable that the interaction of the supernova shock (or any reverse shocks launched later in the explosion) with convective vortices in the progenitor star, or the action of mixing instabilities behind the shock can generate vortices with  nearly radial vorticity and angular momentum vectors
as required for the spin-kick alignment mechanism of \citet{janka_22}. This would imply that one-sided fallback accretion  in explosions of non-rotating stars destroys rather than enforces spin-kick alignment.

There is, however, one exception where eddies with radial vorticity and angular momentum could arise, namely the case of rapid rotation. In rotating shells, a large-scale vorticity field is present, and balance between
gravity, buoyancy, inertial, and pressure forces can
give rise to a quasi-steady state flow structure with
significant non-radial pressure variations. Due to instabilities and mode interactions, such systems often
develop large-scale eddies with radial vorticity as familiar, e.g., from low-Rossby number flow in Earth's atmosphere.
The velocity fluctuations in these eddies tend to be somewhat smaller than the average rotational flow velocity. Hence, the vectorial angular momentum of fallback material will mostly be aligned with the axis of rotation, so that,
as pointed out by \citet{janka_22}, fallback will still lead to spin-kick misalignment much of the time, since the neutron star kick direction appears to be selected randomly
and not in alignment with the rotation axis of the progenitor according to current 3D simulations.

\section{Estimates for Angular Momentum Deposition on the
Neutron Star}
Nevertheless, the concept of one-sided accretion of \citet{janka_22} remains important and potentially useful for constraining supernova physics. If fallback misaligns (rather than aligns) neutron star spins and kicks, this will place limits on the permissible amount of fallback in (typical) core-collapse supernovae if one can estimate the amount of angular momentum contained in accreted vortices, as we shall outline in this section.

Under the assumption that shock-vortex interaction leaves the angular momentum of shocked eddies essentially unchanged (except for some redistribution of angular momentum by waves propagating parallel to the shock surface), the accreted angular momentum can be estimated based on the convective velocities and eddy scales at the pre-collapse stage. The subsequent evolution of shocked vortices that undergo fallback is not straightforward because of vortical-acoustic coupling during the collapse phase
\citep[e.g.,][]{kovalenko_98,lai_00,abdikamalov_19} and Rayleigh-Taylor instability. However, linear theory for perturbations in the accretion flow
\citep[e.g.,][]{kovalenko_98,lai_00,abdikamalov_19} suggests that the
angular momentum contained in transverse motions is approximately conserved
during the infall; in the linear regime, transverse velocity perturbations asymptotically scale as $\delta v_\mathrm{t}/v_r \propto r^{-1/2}$ \citep{lai_00} and hence
as $\delta v_\mathrm{t} \propto r^{-1}$ for free-fall conditions. Rayleigh-Taylor instability will predominantly accelerate plumes in the radial direction and
may not substantially alter the angular momentum contained in transverse velocity perturbations. With all due caveats about the full non-linear evolution of transverse velocity perturbation behind the supernova shock, it appears justified to estimate the accreted angular momentum from transverse perturbations based on pre-collapse conditions.

\subsection{Upper bound for angular momentum deposited by fallback}
Such estimates for the angular momentum in convective eddies at the pre-collapse stage can be based on 1D stellar evolution models implementing mixing-length theory (MLT; \citealt{biermann_32,boehm_58}). MLT provides radial velocity profiles only; the transverse velocities of the biggest eddies (those that span the entire depth of a convective region) near the convective boundaries typically correspond roughly to the maxima of the radial convective 
velocity profiles within a convective zone \citep[cp.\ Figure~10 in][]{mueller_16c}. This places an upper bound of 
$J \sim v_\mathrm{conv} \Delta r \Delta M_\mathrm{eddy}/2$ on the angular momentum of a single eddy (cp.\ Equation~\ref{eq:j_eddy}), which is not too different from the estimates of \citet{gilkis_14}.
If the accreted fallback mass $\Delta M$ originates only from a portion of a pre-colapse eddy
($\Delta M<\Delta M_\mathrm{eddy}$), higher values 
$j \sim v_\mathrm{conv} R$ of the specific angular momentum of the fallback material are conceivable.
In the regime $\Delta M>\Delta M_\mathrm{eddy}$ the angular momentum of several pre-collapse eddies tends to cancel, and may in fact cancel more efficiently than 
estimated by \citet{gilkis_14} for eddies with uncorrelated
angular momenta, but the most optimistic case with the high possible specific angular momentum will be more relevant for the subsequent discussion.

\begin{figure}
    \centering
    \includegraphics[width=\linewidth]{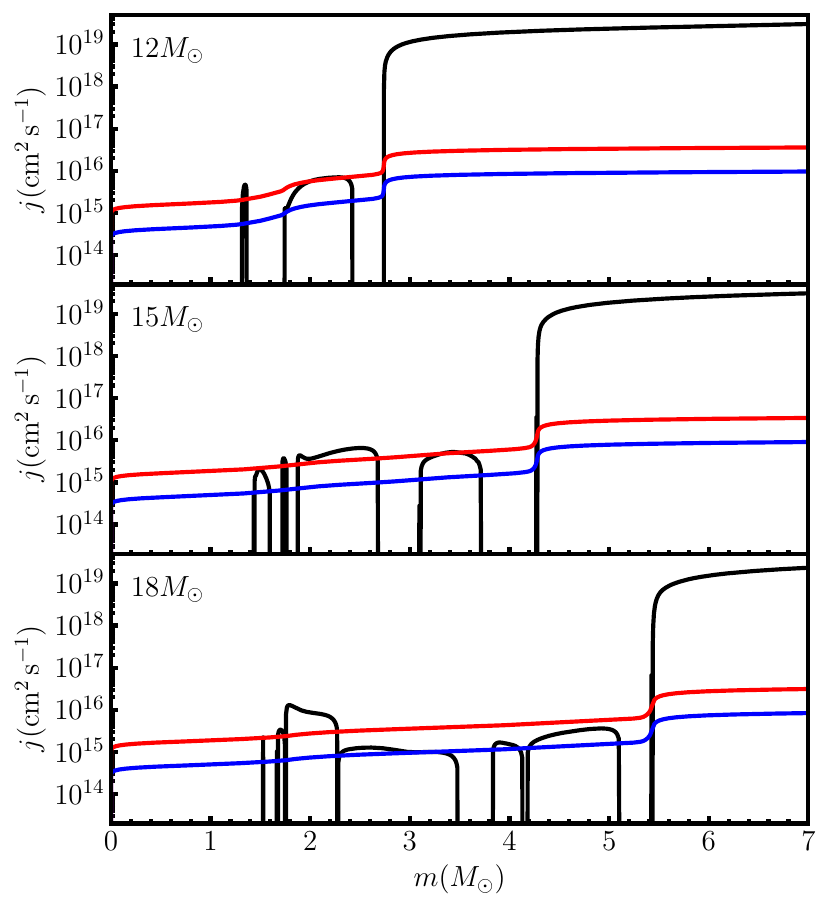}
    \caption{The typical specific angular momentum fluctuations $j=v_\mathrm{conv} R$ associated with convective motions (black) and
    the Keplerian specific angular momentum at the Alfv\'en radius
    (Equation~\ref{eq:spec_j})
    for neutron star magnetic field strengths of $10^{12}\, \mathrm{G}$ (blue) and $10^{14}\, \mathrm{G}$ (red), a fallback mass
    of $\Delta M=0.1 M_\odot$,
    a neutron star mass of $1.4 M_\odot$, and a neutron star radius of $12\,\mathrm{km}$.
    Results are shown for three progenitor models with zero-age main sequence masses of  $12M_\odot$, $15M_\odot$, and $18M_\odot$ from \citet{mueller_16a}.}
    \label{fig:jconv}
\end{figure}

Figure~\ref{fig:jconv} shows the most optimistic value $j \sim v_\mathrm{conv} R$
for the specific angular momentum in pre-collapse eddies for red supergiant progenitors with $12M_\odot$, $15M_\odot$, and $18M_\odot$ from \citet{mueller_16a}. These progenitors are fairly illustrative for the situation in red supergiants at the onset of core collapse in general.
In the hydrogen shell of red supergiants, the specific angular momentum contained in transverse convective motions tends to reach large values of order $10^{18}\texttt{-}10^{19}\, \mathrm{cm}^2\, \mathrm{s}^{-1}$. Even in the inner (oxygen, neon, carbon, helium) shells, the specific angular momentum in convective eddies is still appreciable ($\mathord{\sim}10^{15}\, \mathrm{cm}^2\, \mathrm{s}^{-1}$).
If the specific angular momentum of fallback material were conserved during accretion onto the neutron star, even minimal amounts of fallback from the hydrogen shell could change the neutron star angular momentum appreciably and destroy any pre-existing spin-kick alignment if the neutron star accretes primarily ahead of its direction of travel. Thus, one might surmise that the unavoidable one-sided accretion pointed out by \citet{janka_22} might actually place stringent limits on the amount of fallback matter
from the hydrogen envelopes, and to a lesser extent from inner shells as well. To
impart an angular momentum of
$5\times 10^{47}\, \mathrm{g}\, \mathrm{cm}^2\,\mathrm{s}^{-1}$ perpendicular to the kick direction, which is sufficient to significantly misalign a neutron star with a birth spin period of $\mathord{\sim}20\,\mathrm{ms}$ as for the Crab pulsar \citep{lyne_15} and a typical moment of inertia
of $1.5 \times 10^{45}\, \mathrm{g}\,\mathrm{cm}^2$, fallback of less than
$10^{-1}M_\odot$ in the kick direction from an inner convective shell and of less than $10^{-4}M_\odot$ from the hydrogen envelope would appear sufficient.

However, avoidance of spin-kick misalignment by one-sided accretion does not actually place such stringent limits on the accreted material from the hydrogen envelope. Material with very high specific angular momentum simply cannot be accreted onto the neutron star because of centrifugal support. Unless it manages to lose angular momentum, it will instead bypass the neutron star on an orbit with a large semimajor axis, but will eventually be blown out by the neutron star wind. Less stringent upper bounds for the fallback mass $\Delta M$ may still be formulated after taking into account constraints on the specific angular momentum of accreted matter.

The worst-case misalignment can be estimated by noting that fallback material needs to reach the Alfv\`en radius $R_\mathrm{A}$ of the neutron star in order to be effectively captured and deposit its angular momentum, (provided that the Keplerian velocity at the
Alfv\`en radius is still greater than the corotation
velocity such as to avoid ejection in the propeller 
regime; \citealp{ilianarov_75,piro_11}).
This limits its specific angular momentum to
\begin{equation}
 j=\sqrt{2GM R_\mathrm{A}}
\end{equation}
for parabolic orbits.
Equating the magnetic pressure
(assuming a dipole field with dipole  moment $\mu$) and the ram pressure 
$\rho v^2$, $R_\mathrm{A}$ is determined by
\begin{equation}
\left(\frac{\mu }{R_\mathrm{A}^3}\right)^2=\frac{\dot{M} \sqrt{\frac{2G M}{R_\mathrm{A}}}}{4 \pi  R_\mathrm{A}^2},
\end{equation}
where $\dot M$ is the accretion rate during
the fallback event and the velocity $v$ has been set to the escape velocity.
This yields
\begin{equation}
    R_\mathrm{A}
\sim
    \left(\frac{8 \pi^{2} \mu ^{4}}{G M \dot{M}^2}\right)^{1/7},
\end{equation}
after expressing $\mu$ in terms of the surface dipole field strength $B$ and the
neutron star radius as $\mu= B R^3$.
For a rough estimate of the accretion rate during a fallback event, we can assume
that fallback occurs on a free-fall time scale $\tau_\mathrm{ff}$,
\begin{equation}
    \tau_\mathrm{ff}=\frac{\pi}{2}  \left(\frac{r^3}{G M}\right)^{1/2},
\end{equation}
where $r$ is the initial radial distance of the fallback blob and $M$ is the neutron star mass. Generally, fallback will occur after material has expanded substantially from its initial radial position in the progenitor, but as the final results are not particularly sensitive to $r$ or the timescale of fallback, the initial radial coordinate of fallback material in the progenitor provides a suitable estimate for $\tau_\mathrm{ff}$ for practical purposes.
Approximating the accretion rate
in terms of 
$\tau_\mathrm{ff}$ and the fallback mass $\Delta M$ as
$\dot{M}\sim \Delta M/\tau_\mathrm{ff}$, we find the maximum specific angular momentum of the blob,
\begin{align}
\label{eq:spec_j}
        j&=
    \frac{{2}^{1/14} \pi ^{2/7} B^{2/7} G^{5/14} M^{5/14} r^{3/14} R^{6/7}}{\Delta M^{1/7}}\\
    &=\nonumber
    5.2\times 10^{15}\, \mathrm{cm^2}\,\mathrm{s}^{-1} \times
    \left(\frac{B}{10^{12}\, \mathrm{G}}\right)^{2/7}\times
    \left(\frac{R}{12\, \mathrm{km}}\right)^{6/7}\times\\
    &\left(\frac{M}{1.4 M_\odot}\right)^{5/14}\times
    \left(\frac{r}{10^5\, \mathrm{km}}\right)^{3/14}\times
    \left(\frac{\Delta M}{0.1 M_\odot}\right)^{-1/7}.
    \nonumber
\end{align}
The maximum angular momentum $\Delta J$ is
\begin{align}
\label{eq:delta_j}
    \Delta J&=
    {2}^{1/14} \pi ^{2/7} B^{2/7} G^{5/14} M^{5/14} r^{3/14} R^{6/7}\Delta M^{6/7}\\
    &=\nonumber
    2.8\times 10^{47}\, \mathrm{g}\,\mathrm{cm^2}\,\mathrm{s}^{-1} \times
    \left(\frac{B}{10^{12}\, \mathrm{G}}\right)^{2/7}\times
    \left(\frac{R}{12\, \mathrm{km}}\right)^{6/7}\times\\
    &\left(\frac{M}{1.4 M_\odot}\right)^{5/14}\times
    \left(\frac{r}{10^5\, \mathrm{km}}\right)^{3/14}\times
    \left(\frac{\Delta M}{0.1 M_\odot}\right)^{6/7}.\nonumber
\end{align}
For other purposes, it may be more useful to express
$\Delta J$ directly in terms of the duration $\tau$ of  fallback accretion rather than approximating $\tau$ as the freefall time at a specific radius,
\begin{align}
\label{eq:delta_j2}
    \Delta J&=
    {2}^{3/14} \pi ^{1/7} B^{2/7} G^{3/7} M^{3/7} R^{6/7}\Delta M^{6/7} \tau^{1/7}\\
    &=\nonumber
    3.8\times 10^{47}\, \mathrm{g}\,\mathrm{cm^2}\,\mathrm{s}^{-1} \times
    \left(\frac{B}{10^{12}\, \mathrm{G}}\right)^{2/7}\times
    \left(\frac{R}{12\, \mathrm{km}}\right)^{1/7}\times\\
    &\left(\frac{M}{1.4 M_\odot}\right)^{3/7}\times
    \left(\frac{\Delta M}{0.1 M_\odot}\right)^{6/7}\times
    \left(\frac{\tau}{1\, \mathrm{d}}\right)^{1/7}.
    \nonumber
\end{align}

The maximum specific angular momentum of fallback matter
for $M=1.5M_\odot$, $R=12\, \mathrm{km}$,
and $\Delta M=0.1 M_\odot$ is shown in Figure~\ref{fig:jconv}
for two different values of $10^{12}\, \mathrm{G}$
and $10^{14}\, \mathrm{G}$ for the magnetic field $B$
for the three aforementioned progenitor models. By coincidence, the stochastic angular momentum fluctuations in the inner convective zone shells are close to the maximum sustainable specific angular momentum for fallback accretion. By contrast, fallback material from the hydrogen shell can only be accreted at considerably smaller specific angular momentum than the angular momentum fluctuations present in progenitor stars.

In practice, Equations~(\ref{eq:delta_j},\ref{eq:delta_j2})
can therefore be used to estimate upper bounds for the angular momentum that can be imparted onto a neutron star by fallback and contribute to misalignment. $\Delta J$ can then be compared to the typical angular momentum $J$ of young pulsars. For the Crab pulsar with a magnetic field 
$4\texttt{-}8\times10^{12}\, \mathrm{G}$ \citep{cognard_96,lyutikov_07,kou_15} and a birth spin period of $\mathord{\sim}20\, \mathrm{ms}$ \citep{lyne_15}, and a typical moment of inertia of
$1.5 \times 10^{45}\, \mathrm{g}\, \mathrm{cm}^2$, this amounts to $J\approx 5 \times 10^{47} \, \mathrm{g}\, \mathrm{cm}^2\, \mathrm{s}^{-1}$. Substantial
misalignment ($\Delta J>0.5 J$) would be achieved
for fallback masses $\gtrsim 0.01 M_\odot$ and $R>10^5\, \mathrm{km}$, i.e., whenever fallback is spread over a least a few minutes\footnote{The freefall timescale corresponding to this radius.}. For late fallback on a timescale of $\tau=1\, \mathrm{d}$, about $0.006 M_\odot$
of fallback achieve the same result.

\subsection{Relevance of the bound on accreted angular momentum}
However, Equations~(\ref{eq:delta_j},\ref{eq:delta_j2})
assume optimal conditions for the accreted angular momentum transverse to the kick direction. It is therefore not immediately clear that we can place limits on $\Delta M$ from our estimates for $\Delta J$. However, there are good arguments that the typical angular momentum of accreted fallback matter will not be too far below these optimistic estimates.
In discussing how close the estimates from Equations~(\ref{eq:delta_j},\ref{eq:delta_j2}) are to realistic fallback conditions, several factors need to be considered:
\begin{enumerate}
    \item Uncertainties in the magnitude of the transverse angular momentum fluctuations of matter affected by fallback,
    \item Averaging effects that reduce the total angular momentum of fallback matter,
    \item Orientation effects that reduce the angular momentum perpendicular to the spin direction for a given transverse (i.e., non-radial) angular momentum of fallback matter,
    \item Feedback mechanisms in the vicinity of the neutron star that selectively limit accretion of angular momentum transverse to the kick direction.
\end{enumerate}
It seems unlikely that uncertainties in the transverse angular momentum fluctuations prior to fallback could results in  significantly lower specific angular momentum than suggested by Equation~(\ref{eq:spec_j}). Especially for typical pulsar magnetic field strengths well below the magnetar regime, transverse convective velocities correspond to a specific angular momentum somewhat above the critical value already. Fallback material could, of course, stem from regions that are initially non-convective, but given that a substantial fraction of supernova progenitors is taken up by convective shells, there would need to be a dynamical mechanism that facilitates fallback from non-convective regions in the progenitor. More importantly, even for spherically symmetric progenitor models
long-time 3D supernova simulations \citep{chan_20b,janka_22} show quite robustly that hydrodynamic instabilities that act during the explosion\footnote{This covers both neutrino-driven convection  and the standing accretion shock instability  during the pre-explosion or early explosion phase that imprint asymmetries on the  ejecta initially, as well as mixing instabilities during the propagation of the shock through the envelope.
} generate sufficient transverse velocities for the specific angular momentum in fallback matter 
well above $10^{16}\, \mathrm{cm}^2\,\mathrm{s}^{-1}$, i.e., sufficiently large for disk formation and even for staying outside the Alfv\'en radius.
Thus, one expects that the distribution of specific angular momentum of \emph{accreted} fallback material should extend up to the maximum allowed value and is not skewed towards zero within that range.

It also appears unlikely that variations in the transverse specific angular momentum of fallback matter average out effectively. Numerical simulation \citep{chan_20b} show at most a few phases of spin-up and spin-down even for cases with high fallback masses, indicating that the accreted (vectorial) specific angular momentum is relatively stable and correlated over long time scales. The 3D simulations of \citet{janka_22} also appear compatible with long correlation times in $\mathbf{j}$.
At least for small fallback masses of no more than a few $0.01 M_\odot$, one actually intuitively expects that spin-up by fallback should be determined by only a few spin-up phases with correlated angular momentum. The typical asymmetric structures emerging from the oxygen shell and shell further out during the explosion have low to medium wavenumbers $\ell$ (both because wide convective shells favour large-scale seed structures and because large-scale asymmetries develop generically during the engine phase), and the number of coherent plume structures within a shell will not be excessively high. Even if the fallback material stems from several such plume structures, stochastic cancellations of fallback material from different plumes should not be highly efficient, and the average specific angular momentum of accreted matter will not be reduced by orders of magnitude. Especially for material from the convective hydrogen envelope, where the specific angular momentum in convective seed motion is extremely large, it is therefore improbable that the average specific angular momentum of fallback material can be brought significantly below the critical value at the Alfv\'en radius.

Orientation effects imply that the accreted angular momentum generally has components both parallel and perpendicular to the kick direction. 
\citet{janka_22} demonstrates that the neutron star should accrete from a cone with a half-angle of about $\pi/3$. If, as we have argued, the angular momentum
of the fallback material is mostly perpendicular to its initial radius vector, the accreted angular momentum perpendicular to the kick should exceed a fraction of
$\cot \pi/3=0.58$ of
the component parallel to the kick, but will usually be higher unless some feedback mechanism operating in the vicinity of the neutron star favours accretion of material parallel to the kick (see below). Thus, orientation effects cannot substantially reduce the transverse component of the angular momentum of fallback material either.

Finally, there might be some mechanism that inhibits the accretion of angular momentum perpendicular to the kick direction. This, however, could at best limit accretion to material from near the surface of the accretion cone, i.e., to material that still has considerable angular momentum transverse to the cone as we just discussed. There is no way to simply get rid of this angular momentum component completely once fallback material is approaching the neutron star. It can only be deposited on the neutron star or transferred to other fallback material in the vicinity of the neutron star by some mechanism that separates angular momentum perpendicular and transverse to the kick and then be ejected. However, one should naturally expect that outflows powered by the interaction of neutron stars with accreting matter should extract angular momentum from the the neutron star, i.e., eject material with specific angular momentum aligned to the neutron star, and hence spin it down (cp.\ the case of propeller accretion, \citealp{alpar_01,romanova_04,piro_11}). A mechanism that preferentially powers outflows that carry angular momentum transverse to that of the neutron star may not be impossible, but at present we view this as an unlikely possibility.

Incidentally, misalignment by fallback can, of course, be avoided if the neutron star \emph{is} in the propeller regime and material and gains angular momentum and energy by the interaction
with the magnetosphere. For estimating
limits on the accreted mass compatible
with spin-kick alignment, ejection of fallback material in the propeller regime makes little difference. The possibility that
fallback accretion may sometimes proceed
in the propeller regime \citep{piro_11} does not change the fact that any material that manages to be accreted will add to the neutron star the angular momentum it had prior to interaction with the magnetosphere; hence bounds on the amount of actually accreted material can still be deduced from the condition of limited spin-kick misalignment. It is just that the propeller regime may provide a \emph{physical} mechanism for limiting the amount of accretion even if considerable fallback occurs. However, for large fallback masses, sufficiently early fallback times, and moderately strong neutron star magnetic fields, the propeller regime is avoided and matter that reaches the Alfv\'en radius 
with Keplerian velocity can be accreted. For the aforementioned example of $0.006 M_\odot$
of fallback onto the Crab pulsar on a time scale of a $1\,\mathrm{d}$, the Keplerian specific angular momentum at the Alfv\'en radius is about $3.6\times 10^{16}\,\mathrm{cm}^2\,\mathrm{s}^{-1}$,
whereas the angular momentum for corotation
at the Alfv\'en radius,
$j_\mathrm{c}=2\pi P^{-1} R_\mathrm{A}^2 $,
is only $2.3\times 10^{15}\, \mathrm{cm}^2\,\mathrm{s}^{-1}$. Thus,
for moderately strong neutron star birth magnetic fields, the total amount of material that reaches the Alfv\'en radius 
in the first place likely has to be limited
by the explosion physics to avoid spin-kick misalignment. The propeller mechanism could not stop substantial fallback onto the magnetosphere in such cases, though it may be relevant for limiting accretion at later stages or for stronger neutron star magnetic fields
\citep{piro_11}.

As far as we can see, only one possible loophole for
avoiding spin misalignment in the case of substantial fallback remains. If the kick is \emph{initially} aligned with the rotation axis of the progenitor star, and if rotation is so fast as to push the pre-collapse convection zones into the limit of low Rossby number
(where the rotational velocity exceed the convective velocity), the accreted angular momentum transverse to the kick direction would become negligible. As \citet{janka_22} pointed out, however, this scenario is unlikely in the light of current simulation results that do not show spin-kick alignment even in the case of rapidly rotating progenitor models \citep{powell_20,powell_23}. As long as the initial kick direction is selected randomly, one-sided accretion will
tend to produce spin-kick misalignment in the case of rapid progenitor rotation \citep{janka_22}. Thus, rotation is not likely to alter the findings outlined above. If rotation is fast (with rotation velocities exceeding pre-collapse convective velocities in a particular region), it will at most exacerbate the misalignment because stochastic cancellation of the angular momentum of accreted vortices should become less relevant.

All of these considerations suggest that the specific angular momentum perpendicular to the kick velocity of fallback material
is probably only a factor of a few below our analytic estimates. The estimates from Equations~(\ref{eq:delta_j},\ref{eq:delta_j2}) can therefore indeed be translated into upper limits for the amount of fallback in the formation of typical pulsars. Late-time fallback of not much more than $\mathord{\sim} 10^{-2} M_\odot$ on timescales of a day or more can be tolerated to maintain spin-kick alignment, and perhaps a bit more early on during the explosion. If a neutron star is born with a significantly longer period than $20 \, \mathrm{ms}$ and shows spin-kick alignment, the limit on the amount of fallback accretion becomes more restrictive.
 Interestingly, the fallback masses
of a few $10^{-3} M_\odot$ predicted by the 1D simulations of \citet{ertl_16} for single stars
and  by \citet{ertl_20}
for stripped stars with final helium core masses
below $\mathord{\sim} 6 M_\odot$ are consistent with this constraint.
On the other hand, \citet{ertl_20} predict substantial
fallback of $\mathord{\sim} 0.1 M_\odot$ for some
explosions for massive stripped stars in an island
of explodability with final helium core masses of $\gtrsim 10 M_\odot$. For these, the angular momentum imparted by fallback accretion should be substantial, in line with purely hydrodynamic long-time simulations of fallback in 3D by
\citet{chan_20b}. However, it seems premature to conclude that these explosions of high-mass progenitors should make a sub-population of neutron stars without spin-kick alignment. If a spin-kick alignment mechanism operates early on in these explosions during the ``engine phase'',
the initial angular momentum of the neutron star before
fallback should be unusually high, as it will undergo substantial accretion and strong spin-up during the ``engine phase'' so that spin-tilting requires higher fallback masses.
Moreover, only a small fraction of the neutron star population is expected to be affected by such high amounts of fallback.

\section{Conclusion}
Prompted by recent work of \citet{janka_22} that highlighted the potential impact of one-sided fallback accretion on neutron star birth spins, we further analysed the expected angular momentum carried by vortices in shocked stellar material that may undergo fallback. We argue that one-sided accretion onto moving neutron stars will lead to spin-kick misalignment rather than alignment. Both pre-collapse convective motions and Rayleigh-Taylor instability during the supernova explosion predominantly create eddies with vorticity and angular momentum perpendicular to the radial direction. Accretion of material located in an accretion cone around the neutron star kick direction will therefore mostly add angular momentum perpendicular to the kick direction.
Flattening of eddies by shock compression does not eliminate the transverse angular momentum components. Despite strong strong radial compression, the vorticity and angular momentum of pre-collapse convective eddies remain predominantly non-radial after being run over by the shock, even if the eddies are somewhat deformed and waves may somewhat redistribute angular momentum behind the shock front.

The realisation that fallback will more likely destroy rather than induce spin-kick alignment has implications both for the spin-kick alignment mechanism and for fallback in core-collapse supernovae. First, the observed spin-kick alignment obviously still remains unexplained. If the mechanism involves (magneto-)hydrodynamic torques by accretion downflows, it likely has to operate early on during the explosion. Second, the tendency of fallback to misalign neutron star spins and kicks suggests limits on the typical amount of fallback that can be tolerated in supernova explosions. 

In estimating spin-kick misalignment from fallback, one needs to take into account that accreted material has to come within about the Alfv\'en radius of the neutron star to effectively deposit its angular momentum. Based on this notion, we derived scaling laws for the
maximum specific and total angular momentum of fallback 
(Equations~\ref{eq:spec_j}--\ref{eq:delta_j2}). 
Estimates of the angular momentum contained in pre-collapse convective eddies as well as 3D simulations of fallback \citep{chan_20b,janka_22} suggest that the accreted material easily reaches or exceeds the Keplerian angular momentum
at the Alfv\'en radius for typical neutron star birth magnetic field strengths, and that stochastic variations  in  the angular momentum direction of accreted matter should not push the estimates from Equations~(\ref{eq:spec_j}--\ref{eq:delta_j2}) down substantially. Equations~(\ref{eq:spec_j}--\ref{eq:delta_j2}) imply that young neutron stars should not accrete more than $\mathord{\sim}10^{-2}M_\odot$ by fallback during the first day(s) of an explosion; otherwise the accreted angular momentum would be sufficient to misalign a neutron star with a spin period of $20\, \mathrm{ms}$. For neutron stars with longer birth periods, the limit becomes more stringent. Pre-collapse rotation is not likely to weaken these upper bounds. The fallback masses currently predicted by parameterised 1D supernova explosion models \citep{ertl_16,ertl_20}
are essentially compatible with these limits.

It is of course desirable to put the analytic estimates for neutron star spin-up and spin-kick misalignment by fallback on a firmer footing with the help of multi-dimensional simulations. The work by \citet{janka_22} and our current study highlight, however, that there are non-trivial requirements for correctly tracking spin-up by fallback in such simulations. The motion of the neutron star, the three-dimensional structure of \emph{all} the convective shells in the progenitor, and the coupling of fallback accretion streams with the neutron star magnetosphere may all turn out to be relevant. Whether a rigorous treatment of all of these effects in simulations is possible and can add yet another twist to the problem of spin-kick alignment remains to be seen.

\section*{Acknowledgements}
I acknowledge helpful discussions with I.~Mandel. This work was supported by ARC Future Fellowship FT160100035 and by the Australian Research Council (ARC) Centre of Excellence (CoE) for Gravitational Wave Discovery (OzGrav) project number CE170100004.

\section*{Data Availability}

The data from our simulations will be made available upon reasonable requests made to the author. 

\bibliographystyle{mnras}
\bibliography{paper}
\bsp
\label{lastpage}

\end{document}